# Investigation of the Impacts of COVID-19 on the Electricity Consumption of a University Dormitory Using Weather Normalization

**Zhihong Pang**
*Student Member ASHRAE*

**Fan Feng**
*Student Member ASHRAE*

**Zheng O'Neill, PhD, PE**
*Member ASHRAE*

## ABSTRACT

*This study investigated the impacts of the COVID-19 pandemic on the electricity consumption of a university dormitory building in the southern U.S. The historical electricity consumption data of this university dormitory building and weather data of an on-campus weather station, which were collected from January 1$^{st}$, 2017 to July 31$^{st}$, 2020, were used for analysis. Four inverse data-driven prediction models, i.e., Artificial Neural Network, Long Short-Term Memory Recurrent Neural Network, eXtreme Gradient Boosting, and Light Gradient Boosting Machine, were exploited to account for the influence of the weather conditions. The results suggested that the total electricity consumption of the objective building decreased by nearly 41% (about 276,000 kWh (942 MMBtu)) compared with the prediction value during the campus shutdown due to the COVID-19. Besides, the daily load ratio (DLR) varied significantly as well. In general, the DLR decreased gradually from 80% to nearly 40% in the second half of March 2020, maintained on a relatively stable level between 30% to 60% in April, May, and June 2020, and then slowly recovered to 80% of the normal capacity in July 2020.*

## INTRODUCTION

### Background

Since its outbreak in the early beginning of 2020, the pandemic of the coronavirus disease 2019 (COVID-19) caused by the Severe Acute Respiratory Syndrome Coronavirus 2 (SARS-CoV-2) has affected over 200 countries/regions in the world (World Health Organization 2020). According to the U.S. Centers for Disease Control and Prevention (CDC), as of October 31$^{st}$, 2020, there has been a cumulative total of nearly nine million confirmed COVID-19 cases being reported in the U.S., of which over two hundred thousand deaths were included (U.S. CDC 2020).

Amid this exigent situation, as of May 13$^{th}$, 2020, 41 U.S. states have issued stay-at-home orders forcing the closure of all non-essential business properties in response to the substantial threat of COVID-19 to public health (National Conference of State Legislatures 2020a). As the COVID-19 outbreak continued to sweep through the country, the national electricity usage dropped dramatically as businesses shuttered and people and students hunkered down in their homes (Hsu 2020). A visualization of the U.S. National Aeronautics and Space Administration (NASA) "Black Marble" satellite data showed that the nighttime lights of New York City dimmed by 40 percent between February and April 2020 (Ruan et al. 2020). As estimated by the U.S. Energy Information Administration (EIA), the recent business shutdowns related to COVID-19 mitigation efforts have caused a 9%–13% reduction to the daily weekday electricity demand in the central region of the U.S. in March and April 2020 compared with the expected demand (weather normalization was considered) (U.S. EIA 2020).

Besides business, governors and legislatures have also called for the statewide closure of at least 124,000 public schools in 48 states and every U.S. territory (National conference of State Legislatures 2020c). Also, by the middle of March 2020, more than 1,100 U.S. colleges and universities in all 50 states have announced the cancellation of classes and/or a shift to

**Zhihong Pang** and **Fan Feng** are PhD students, and **Zheng O'Neill** is an Associate Professor at J. Mike Walker 66' Department of Mechanical Engineering in Texas A&M University, College Station, TX.

online-only instruction for the remaining Spring semester of 2020 (National Conference of State Legislatures 2020b). This transition, on the one hand, reduced the infection risk of the college students in high-density classroom settings; on the other hand, however, also greatly changed the campus electricity usage profile since most of the students have travelled back to home and many university facilities (e.g., dormitories and cafeterias) were closed or operated with a reduced capacity.

Data from the City University of New York Brooklyn College showed that compared with the pre-pandemic period, its weekly electricity consumption on campus in April 2020 reduced by nearly 21% to 36% in terms of peak load, and 19% to 22% in terms of total load (Tomkiewicz 2020). Data revealed by Cornell University suggested that compared with the pre-pandemic week of March $9^{th}$ to March $15^{th}$, 2020, the peak electricity load nearly dropped by 28%, which was nearly eight megawatts, in the post-pandemic week of March $23^{rd}$ to $29^{th}$ (Cornell Campus Sustainability Office 2020). Similarly, the energy bill of Western Washington University (WWU) showed that the campus consumed about 6349 thousand kWh (21662 MBtu) of electricity from April to July 2020, which was 30% less than the same period in 2019 (WWU 2020).

**Weather Normalization**

A clear understanding of the building's electricity profile is a necessity and premise to secure and optimize the smooth and resilient operation of the power grid. With the concept of Grid-interactive Efficient Buildings (GEBs) gaining wide recognition and acceptance from the academia and industry (U.S. Department of Energy 2019), knowing how the building load profile was re-shaped under abnormal circumstances (e.g., the COVID-19 pandemic) serves as a valuable resource to enhance the resiliency, reliability, and sustainability of grid.

As demonstrated in the literature review, some universities have disclosed their electricity consumption data before and after the outbreak of the COVID-19 pandemic. Although these data could help reveal how the campus closure has reshaped the building load pattern, a weather normalization should be performed to eliminate the influences of weather variation on energy consumption before the data could be used for load comparison. Dynamic variations in weather conditions directly pose an influence on energy consumption inside the building system (Al-Hadhrami 2013). Aimed at weakening or even eliminating the impacts of different weather conditions, the weather normalization techniques permit a fair comparison of building energy consumptions in multiple different years (Eto 1988).

There have been several weather normalizations methods explored in existing studies, e.g., the degree day methods (Quayle and Diaz 1980), extended forms of degree day methods (Cesaraccio et al. 2001; Schoenau and Kehrig 1990), ASHRAE change-point models (Kissock et al. 2001), Energy Signature methods (Santamouris 2010), Climate Severity Index method (Markus 1982), energy simulation-based methods (Wang et al. 2017), etc. However, they are either too simplified to generate accurate weather normalized results, or too complicated for practitioners to perform (e.g., there is a need of developing a building energy model for simulations), hence are not the most optimal candidates for weather normalization of university electricity consumption, considering that there are usually multiple buildings on campus and the fuel source and the mechanical system and occupant loads may vary significantly with buildings.

Data-driven machine learning (ML) algorithms, e.g., support vector machine, is a feasible and efficient solution to predict building energy consumption using the inputs such as weather conditions, load conditions, and building occupancy (Zhao and Magoulès 2012). Further, with substantial progress of deep learning (DL), the data-driven/inverse model-based building energy consumption prediction has kept pushing the boundary for higher accuracy (Neto and Fiorelli 2008). Given the many successful case studies, machine learning and deep learning could serve as an excellent candidate for weather normalization.

**Objective**

This study aims to investigate how the COVID-19 pandemic has impacted the electricity load profile of a university building using electricity data and weather measurements. The results are expected to provide insight and guidance on navigating the resilience and operation optimization of the grid and buildings to prepare for the next public health emergency should there be any in the future.

This paper is structured as follows. Firstly, the background of the COVID-19 pandemic, its impacts on building electricity consumption and grid operation, as well as the prevailing methods for weather normalization of energy consumption was presented. Then, the four ML/DL models used in this study for building energy prediction and weather normalization, i.e., Artificial Neural Network (ANN), Long Short-Term Memory (LSTM) network, eXtreme Gradient Boosting (XGBoost), and Light Gradient Boosted Machine (LightGBM), were briefly elaborated in terms of principles, advantages, and disadvantages.

Next, the object building, submetering system, and weather station used in this study were introduced. The collected energy data were normalized based on the aforementioned four models. Finally, the results were presented and discussed.

## METHODOLOGY

As mentioned previously, four prediction models were selected for building energy prediction and weather normalization in this study. The methodology of ML/DL based weather normalization for building electricity was illustrated in Figure 1. There are three steps in general:

- Train the ML/DL model with the pre-pandemic building electricity consumption data and weather measurement;
- Predict the building electricity consumption during the pandemic using the trained ML/DL model and the weather measurement during the pandemic;
- Compare the predicted electricity consumption data against the measured electricity consumption data and analyze the results.

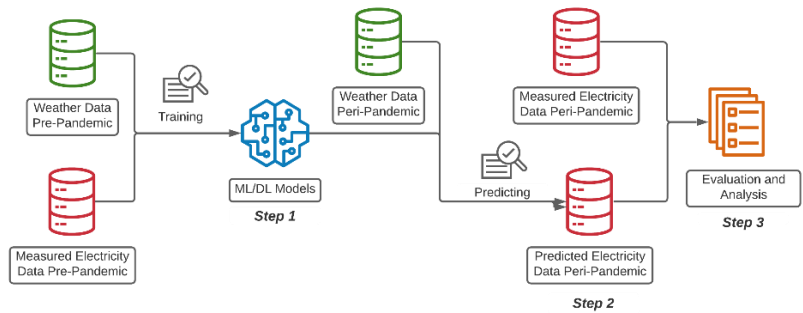

**Figure 1**   The methodology of ML/DL based weather normalization for building electricity.

The principles, advantages, and disadvantages of the four selected ML/DL models are briefly introduced below.

### Artificial Neural Network

Since the first McCulloch-Pitts neural model was presented in 1943, hundreds of different varieties of ANN models have been developed and optimized for various applications like pattern recognition, prediction, optimization, associative memory, and controls. The concepts of the ANN were inspired by the biological neural networks that constitute animal brains (Jain et al. 1996). The structure of a most common and simple feed-forward is illustrated in Figure 2 (a) (Pang et al. 2020). As shown, a typical feed-forward ANN model usually has three layers, i.e., the input layer which receives inputs from outside, the hidden layer which conducts matric calculations to generate intermediate outputs, and the output layer which outputs final prediction results for the analysis (Neto and Fiorelli 2008). It should be noted that the number of hidden layers is not necessarily one; different ANN configurations could have none, one, or multiple hidden layers depending on the data and purposes.

The two competitive advantages of ANN against other traditional prediction models, e.g., linear regression, are that firstly, it could well handle the non-linearity of the data, and secondly, an ANN could learn by itself (Ahmad et al. 2014). However, ANN also suffers from several drawbacks. Some studies argued that the benefits of the ANN model over other data mining techniques are marginal (Ahmad et al. 2017). Besides, it usually takes a long time and large computational power to train an ANN model, especially when there are multiple hidden layers.

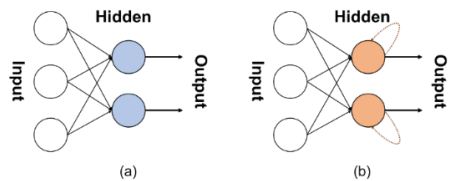

**Figure 2**   The methodologies of (a) simple feed-forward ANN, and (b) RNN.

### Long short-term memory network

RNN is a class of ANN where connections between units form a directed cycle. The key idea behind RNN is to utilize sequential information. In a traditional neural network, all inputs (and outputs) are assumed to be independent of each other. However, for many applications, this assumption is not always true. RNNs are called recurrent because they repeat the same task for every element of a sequence, with the output being dependent on the previous computations, as illustrated in Figure 2 (b). In other words, RNNs have a memory that captures the information about what has been calculated so far.

LSTM is a special kind of recurrent neural network (RNN). It was designed to solve the limitations of traditional RNN models, e.g., the gradient vanishing and exploding problems. The core of the LSTM model is a memory cell, of which the state may vary over time. Gated nonlinear units are used to decide which information should be thrown away from or stored in the cell state. This special structure gives the LSTM model some advantages, e.g., it does not have to suffer from the optimization hurdles which trouble the simple recurrent networks (SRNs) (Greff et al. 2016).

The advantages of LSTM over ANN and SRN are obvious: it could model time series data in which each sample is dependent on previous ones just as good as a normal RNN, but it could also mitigate the gradient vanishing issue of SRNs brought with long memories. The disadvantages of LSTM are obvious too: it could not eliminate the issue of gradient vanishing, and it takes even more computational power than ANN (LeCun et al. 2015).

**eXtreme Gradient Boosting**

The XGBoost algorithm is developed based on the gradient boosting tree, which is also known as gradient tree boosting, or gradient boosting machine (GBM) (Chen and Guestrin 2016). Boosting refers to an ensemble technique that creates multiple models and then combines the models to an ensemble, which produces more accurate prediction results than single leaners. The models are sequentially added to the combination until no further improvements can be made (Zhou 2009).

The XGBoost algorithm has been widely used by many practitioners, and its performance in prediction accuracy is widely recognized and appreciated. Besides, XGBoost incorporates a regularized model to prevent the overfitting issue. The regularization of XGBoost resembles previous work on a regularized greedy forest, but the objective and algorithm are simplified for scalability and parallel calculations (Chen and Guestrin 2016). Despite the pros, one big issue with XGBoost is that it tends to consume a lot of computational power and time to train the model. Also, some studies showed that XGBoost may still suffer from the overfitting issue in some scenarios despite the regularized model. Some practitioners commented that tuning the hyperparameters of the XGBoost model could be a tricky job.

**Light Gradient Boosting Machine**

Similar to XGBoost, the LightGBM was also developed based on GBM. However, to improve the efficiency and scalability in the circumstance of high feature dimensionality and large data size, two unique techniques are applied by LightGBM. Firstly, with Gradient-based One-Side Sampling (GOSS), a significant proportion of data instances with small gradients is excluded. This could help obtain more accurate estimations of the information gain with much smaller data size. Secondly, with Exclusive Feature Bundling, the mutually exclusive features are excluded, so the number of features is reduced. These two techniques together help LightGBM achieve good accuracy with fast training (Ke et al. 2017).

One issue with LightGBM is that this method is only recommended for big datasets since it is very sensitive to overfitting. Hence, it may underperform when being applied to small datasets.

**CASE STUDY**

**Building Facts**

The building in this study is a Dormitory building at the University of Alabama, Tuscaloosa, AL. This building has four stories above the ground, and each floor holds 28 suites which are served by a split direct expansion (DX) air handling unit (AHU) with a hot water coil. The hot water is supplied by a central plant. With a total floor area of 12,150 m$^2$ (130,782 ft$^2$), this dormitory building is designed to provide on-campus accommodations for up to 368 university students and faculty members with one-, two-, and four-bedroom suites. The outlook of this building in Google Maps was presented in Figure 3 (a).

The total electric energy consumption, covering lighting, plug load, and heating ventilating and air-conditioning (HVAC) load for each floor, is sub-metered using the WattNode submetering system. The data has been sub-metered with a sampling frequency of 2 minutes since July 2014. The sub-metered data is automatically pushed to an online server every 10 minutes for convenient retrieval and archival purpose.

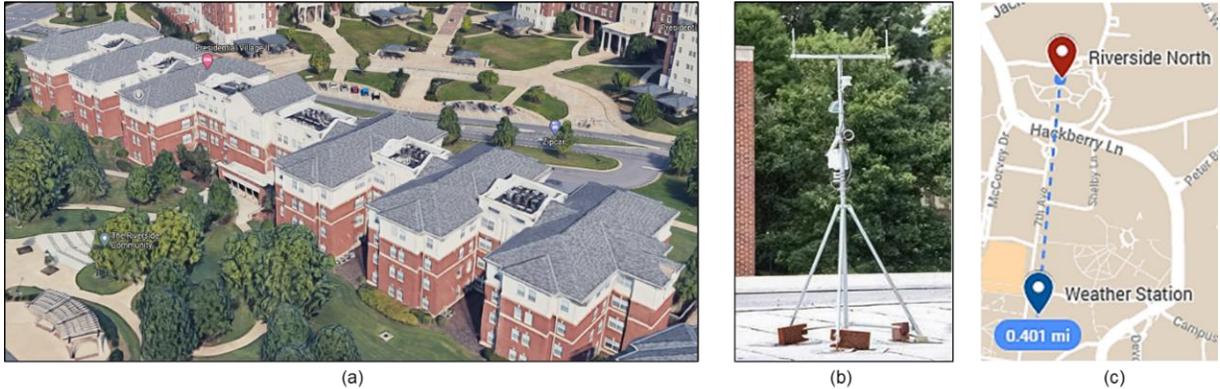

**Figure 3**   (a) The outlook of the Dormitory in Google Maps; (b) The photo of the onsite weather station; (c) The location of the dormitory and weather station in Google Maps.

**Weather Station**

The weather data used in this case study was collected from an on-site weather station, as shown in Figure 3 (b). This weather station is about 650 m (2132 ft) to the South of the study building, as shown in Figure 3 (c). The campus is located in the southeast of the United States and with a coordinate of 33°12' N and 87°32' W. The local climate is blessed with abundant solar resources in the summer. The meteorological data being collected by the weather station includes global solar radiation, outdoor air-dry bulb temperature, relative humidity, dew-point temperature, wind speed, and wind direction.

Similar to the sub-metering system, the data of weather station had been recorded with a sampling frequency of 2 minutes since July 2014 and is pushed to the online server every 10 minutes.

**Data Pre-Processing**

The electricity and weather data retrieved from the server were scrutinized and pre-processed to ensure data quality. The missing data were replaced by the mean value of the previous and next data points. Besides, the raw data of the high sampling frequency were converted to the daily data to facilitate the prediction.

The measurements of dry-bulb temperature and solar radiation between January 1st, 2017, and July 31st, 2020 were presented in Figure 4. The daily electricity consumption of the same timespan was presented in Figure 5. As shown, while the temperature and solar radiation stayed on a relatively stable level in 2020 compared with the same period in the previous three years, the building electricity consumption reduced significantly between April 1st, 2020 to July 31st, 2020.

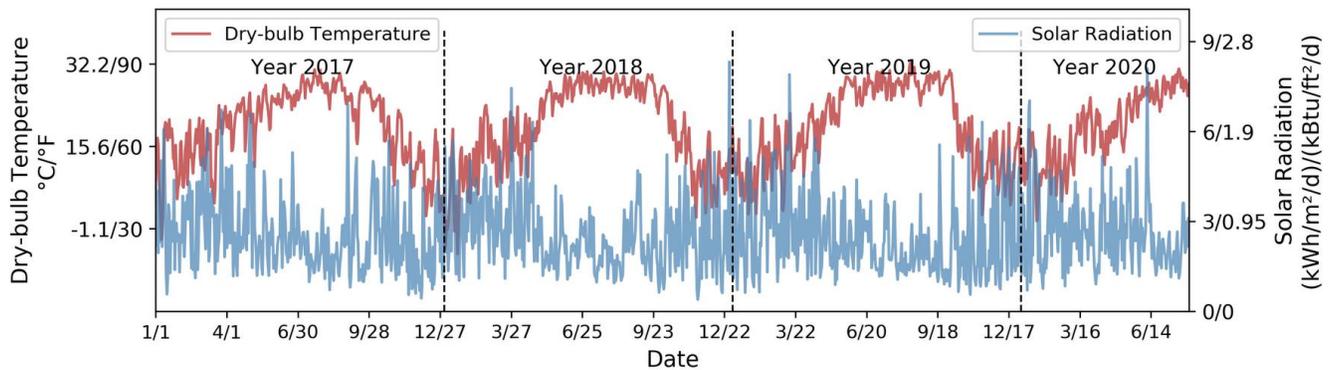

**Figure 4**   Measurement data of dry-bulb temperature and solar radiation between January 1st, 2017 and July 31st, 2020.

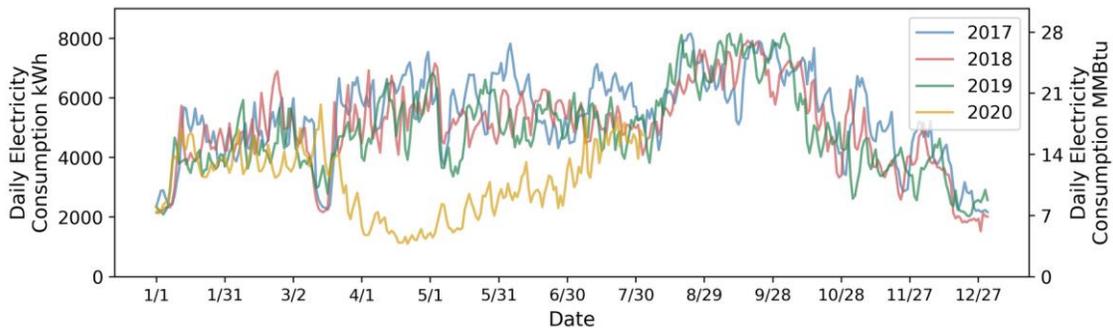

**Figure 5**   Measurement data of daily electric energy consumption between January 1st, 2017 and July 31st, 2020.

In this study, the data of 2017 and 2018 were selected as the training data, and the data of 2019 were chosen for testing the model performance. Considering that the University of Alabama announced the transition to the online class on March 12th (The Crimson White 2020), the period of March 12th, 2020 and July 31st, 2020 was selected as the COVID-19 pandemic period for investigation.

## RESULTS

### Prediction Accuracy

The measurement and prediction results of the testing data (i.e., the whole-building electricity consumption in 2019) were presented in Figure 6. The four subplots correspond to the four selected prediction models respectively, with the red line representing the measurement data and the blue line representing the prediction results.

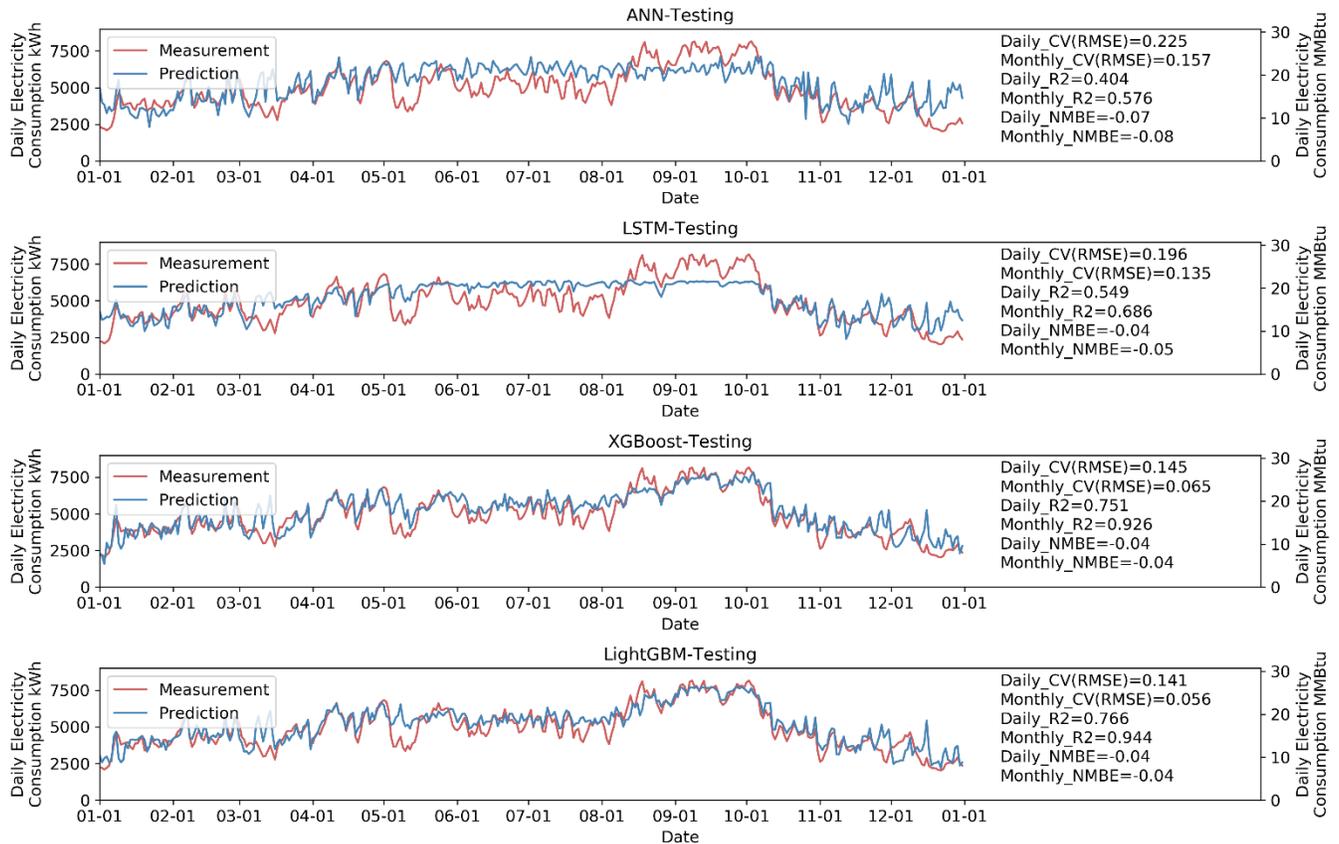

**Figure 6**   The measurement and prediction results of the testing data (January 1st, 2019 to December 31st, 2019).

To quantify the accuracy of the models, the Coefficient of Variation of Root Mean Squared Error (CV(RMSE)), R-squared values ($R^2$), and Normalized Mean Bias Error (NMBE) were computed and annotated in the plots. CV(RMSE), $R^2$, and NMBE are three key performance indexes (KPI) that have been frequently used to measure the differences between values predicted by a model and the values observed. Their calculation is defined in Eq. (1) to (4), where $x_i$ is the actual electric energy consumption, $\hat{x}_i$ is the predicted electricity energy consumption, $N$ is the number of data points, $\bar{x}$ is the mean value of the actual electric energy consumption, and $p$ is the number of parameters in the regression model. The KPIs were computed both for daily data and monthly data.

$$RMSE = \sqrt{\frac{\sum_{i=1}^{N}(x_i - \hat{x}_i)^2}{N}} \tag{1}$$

$$CV(RMSE) = RMSE/\bar{x} \tag{2}$$

$$R^2 = 1 - \frac{\sum_i^N (x_i - \hat{x}_i)^2}{\sum_i^N (x_i - \bar{x})^2} \tag{3}$$

$$NMBE = \frac{\sum_{i=1}^{N}(x_i - \hat{x}_i)}{(N-p)*\bar{x}} \tag{4}$$

As required by ASHRAE Standard 14 (ASHRAE 2014), a building energy model could be considered acceptable with a monthly and daily CV(RMSE) smaller than 15% and 22% respectively, and a monthly and daily NMBE value smaller than 5% and 7% respectively. As shown in Figure 6, the LightGBM model achieved the highest prediction accuracy in terms of both CV(RMSE), $R^2$, and NMBE, then followed by XGBoost, LSTM, and ANN. In this study, the LightGBM and XGBoost methods were finally selected for evaluating the electricity consumption during COVID-19.

## Electricity Consumption During COVID-19

The measurement and prediction results of the whole-building electricity consumption during the COVID-19 pandemic were presented in Figure 7. The results showed that the actual electricity consumption started to significantly deviate from the prediction value after March 12th. This is because though the first U.S. COVID-19 case was confirmed on February 26th, 2020 (Holshue et al. 2020), the campus-wide transition to online classes was announced on March 12th, 2002 (The Crimson White 2020). Generally, both XGBoost and LightGBM output much higher precited electricity consumption results compared with the actual measurements.

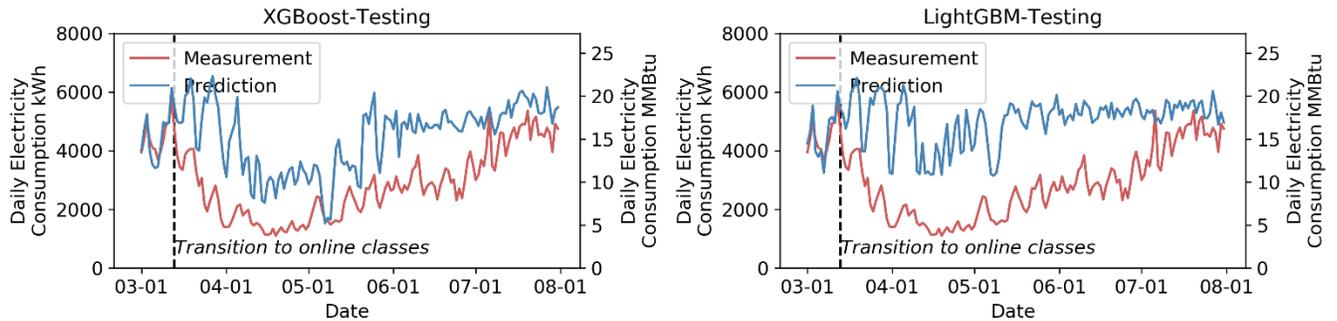

**Figure 7** The measurement and prediction results of whole-building electricity consumption during the COVID-19 pandemic (March 1st, 2020 to July 31st, 2020).

To further quantify the impacts of the campus shutdown due to the pandemic on electricity consumption, the cumulative electricity consumption of measurements and predictions in the investigation period were presented in Figure 8. The blue stacks represent the cumulative electricity consumption of the prediction results, and the red stacks represent that of the actual measurements. Besides, the daily load ratio (DLR), which is defined as the ratio of actual daily electricity consumption by the prediction value, was plotted to show the daily fluctuations as well. Its value is indicated by the gray dashed-dotted lines.

As shown in the figure, the XGBoost model suggested that there was a 37% reduction in total electricity consumption from March 12th to July 31st, 2020, which was about 235,000 kWh (801 MMBtu), while the LightGBM model suggested a 44% reduction, which was about 317,000 kWh (1,082 MMBtu). If we take the mean value of the prediction results from XGBoost and LightGBM as the final prediction result for these four and a half months, this would lead to a total reduction of 276,000 kWh (942 MMBtu) in electricity consumption, which is nearly a 15% of the annual electricity consumption of this building in

2019.

The DLR varied a lot in this period, as suggested by the black lines of Figure 8, partly due to the impacts of the occupant behaviors, which were not taken into account in the current models because of a lack of information. In general, the DLR decreased gradually from 80% to nearly 40% in the second half of March, maintained on a relatively stable level between 30% to 60% in April, May, and June, and then slowly recovered to 80% of the normal capacity in July. This pattern could probably happen due to two reasons. Firstly, the campus was only planned to be closed for one week in the first announcement of transition to online classes considering the ambiguous uncertainties of COVID-19 in its early stage; hence some students might choose to stay on campus and wait for reopening until late March when the University decided to close the campus for the rest of the Spring semester. Secondly, the students began to return to campus successively in July for the Fall semester, so the DLR climbed up a little bit.

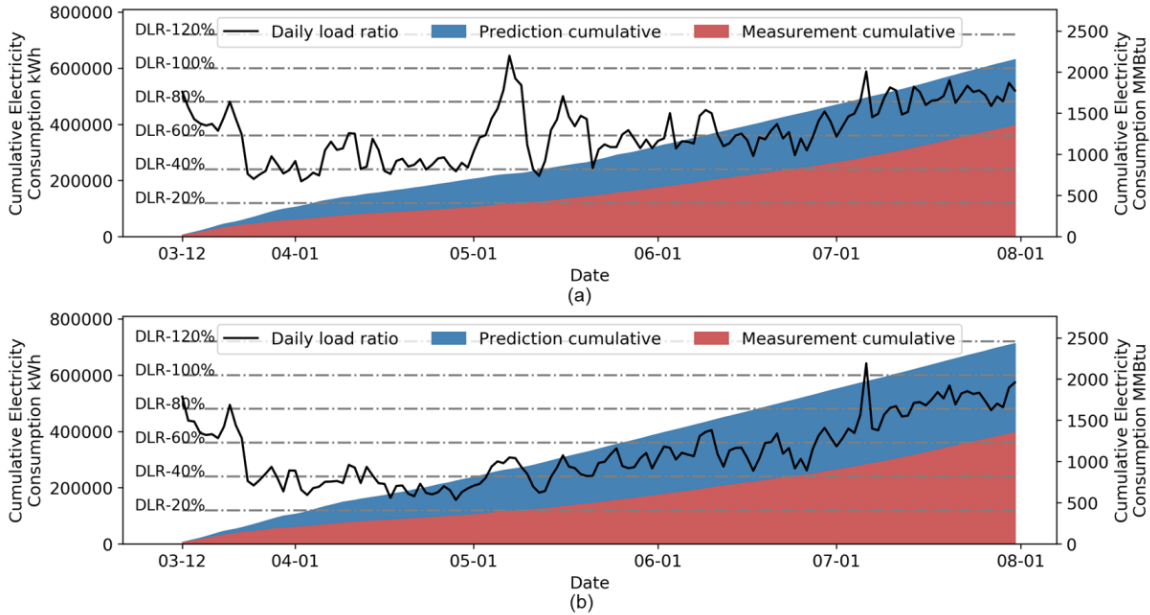

**Figure 8**    The daily load ratio and cumulative electricity consumption of measurements and predictions in the investigation period for (a) XGBoost and (b) LightGBM.

## CONCLUSIONS AND FUTURE WORK

This study investigated the impacts of the COVID-19 pandemic on the electricity consumption of a university dormitory building in the southern U.S. The historical electricity consumption data and meteorological data of an on-campus weather station collected from January 1st, 2017 to July 31st, 2020 were used for the analysis. Four prediction models, i.e., Artificial Neural Network, Long Short-Term Memory Recurrent Neural Network, eXtreme Gradient Boosting, and Light Gradient Boosting Machine, were developed to account for the influence of the weather conditions.

The results suggested that from March 31st to July 31st, 2020, when the campus was closed due to COVID-19, the total electricity consumption of the study building decreased by nearly 41% (about 276,000 kWh (942 MMBtu)) compared with the prediction value with the impacts of the weather being removed. The reduction in these four and a half months equals 15% of the annual electricity consumption of this building in 2019. Besides, the daily load ratio varied significantly as well. In general, the DLR decreased gradually from 80% to nearly 40% in the second half of March, maintained on a relatively stable level between 30% to 60% in April, May, and June, and then slowly recovered to 80% of the normal capacity in July.

Although a relatively high prediction accuracy in terms of CV(RMSE), $R^2$, and NMBE was achieved by two of the four selected models, the accuracy could still be further improved by incorporating more information such as the occupancy level of the building. In the future, the real-time occupancy data extracted from social media, e.g., Facebook and Twitter, could be added to the models to further enhance the prediction accuracy. Besides, considering that multiple universities have disclosed the campus electricity bill to the public for research purposes, the database of this study could be extended to different universities in different climate zones to reveal more insights.